\documentclass[aps,showpacs,twocolumn,
superscriptaddress]{revtex4}
\usepackage{graphicx}
\usepackage{dcolumn}
\usepackage{bm}
\usepackage{color}
\usepackage[normalem]{ulem} 
\usepackage[dvipsnames]{xcolor} 
\usepackage{hyperref}
\hypersetup{
  colorlinks=true,        
  linkcolor=blue,         
  citecolor=cyan,         
}
\usepackage{mathrsfs}
\usepackage[utf8]{inputenc}
\usepackage{mathtools}
\usepackage{doi}
\usepackage{amsmath}
\usepackage{amssymb}

\begin{document}

\title{
Spinning magnetized particles orbiting magnetized Schwarzschild black holes}

\author{Farrux Abdulxamidov} \email{farrux@astrin.uz}
\affiliation{School of Mathematics and Natural Sciences, New Uzbekistan University, Mustaqillik Ave. 54, Tashkent 100007, Uzbekistan}
\affiliation{Ulugh Beg Astronomical Institute, Astronomy St 33, Tashkent 100052, Uzbekistan} \affiliation{Institute of Nuclear Physics, Ulugbek 1, Tashkent 100214, Uzbekistan}

\author{Javlon Rayimbaev}
\email{javlon@astrin.uz}
\affiliation{Institute of Fundamental and Applied Research, National Research University TIIAME, Kori Niyoziy 39, Tashkent 100000, Uzbekistan}
\affiliation{Akfa University, Kichik Halqa Yuli Street 17,  Tashkent 100095, Uzbekistan}
\affiliation{National University of Uzbekistan, Tashkent 100174, Uzbekistan}

\author{Ahmadjon~Abdujabbarov}
\email{ahmadjon@astrin.uz}
\affiliation{Ulugh Beg Astronomical Institute, Astronomy St 33, Tashkent 100052, Uzbekistan} 
\affiliation{National University of Uzbekistan, Tashkent 100174, Uzbekistan}

\author{Zden\v ek Stuchl\'ik}
\email{zdenek.stuchlik@fpf.slu.cz}
\affiliation{Research Centre for Theoretical Physics and Astrophysics, Institute of Physics, Silesian University in Opava, Bezru\v covo n\' am. 13, CZ-74601 Opava, Czech Republic }

\date{\today}

\begin{abstract}
A way to test electromagnetic field and spacetime properties around black holes is by considering the dynamics of test particles. In fact, in real astrophysical scenarios, it is hard to determine spacetime geometry which is dominating due to degeneracy gravitational effects in parameters of gravity theories. In this work, we study for the first time the dynamics of spinning particles that have magnetic dipole moments around Schwarzschild black holes immersed in an external asymptotically uniform magnetic field using the Mathisson-Papapetrou-Dixon (MPD) equation. There are two combined interactions: gravitational interaction between the spin of the particle and (electro)magnetic interaction between the external magnetic field and the magnetic dipole moment of the particle to be taken into account. First, we derive the effective potential of the test spinning magnetized particles in motion around the black hole. We also study the combined effects of spin and magnetic interactions on innermost stable circular orbits (ISCOs), the energy, and angular momentum of the particles at ISCO together with superluminal bounds. We investigated the collision of the particles and evaluated the center-of-mass energy in the collisions. Finally, we consider various cases in which neutron stars and rotating stellar mass black holes can be treated as spinning magnetized particles, evaluating the effects of the spin and magnetic moment of objects around supermassive and intermediate-mass black holes. It is also found that magnetic interaction effects are much larger than spin ones in the case of a neutron star orbiting a supermassive mass black hole, while for the case of a neutron star and intermediate-mass black hole system, the effects are comparable where the magnetic field value is larger than 20 G for typical neutron stars {and this value for the system with rotating stellar mass black holes is about 280 G}.

\end{abstract}

\maketitle

\section{Introduction}

Astrophysical point of view, considering the dynamics of non-zero spin ($s\neq0$) particles around black holes may help to deeply understand astrophysical scenarios of rotating neutron stars, in particular, millisecond pulsars around the supermassive black hole Sagittarius A$^*$ (Sgr A$^*$). Current observations of GRAVITY collaboration on searching pulsars around the Sgr A$^*$, unluckily, show that no pulsar in the close environment of the object. One can explain it by the scattering of radio waves in the dense plasma medium around Sgr A$^*$ or the magnetic interaction between the dipole moment of neutron stars and the external magnetic field.

Since neutron stars and white dwarfs are highly magnetized objects with high magnetic dipole moment $\mu \sim B_s R^3$. In this sense, in studies of the objects as a test particle dynamics, it is important to consider the interaction between the magnetic field in the black hole environment and the magnetic dipole moment of the magnetized objects. The first consideration of magnetized particles' dynamics around a Schwarzschild black hole in the presence of an external test asymptotically uniform magnetic field is studied in Ref.\cite{deFelice}.

Later, studies of magnetized particles' motion around rotating Kerr spacetime, magnetized and magnetically charged black holes in gravity theories have been developed in Refs.\cite{defelice2004,2020PhRvD.102f4052T,2022MPLA...3750220R,2020PhRvD.101l4039V,Abdujabbarov2020Galaxy,Rayimbaev2021NuPhB,2020PhRvD.102h4016R,2020PDU....3000715A,2021IJMPD..3050019R,2021EPJC...81..269N,Bokhari2020PRD,Juraeva2021EPJC,Dilshod2023EPJP,Rayimbaev2023Univ,AqilaR2023EPJP,Zahid2022EPJC,Nuriddin2022IJMPD} and it's been found that there is a limit for magnetic interaction parameters in which the ISCO goes to infinity or disappears.

Studies of the spinning particles need to consider a different kind of symmetry in the behaviour of interaction between the gravitational field and spin parameter depending on the sign of the spin $s$, like the particle's angular momentum,  $\mathcal{L}$, with two different configurations which can be defined by the two symmetries where the effective potential take the same behaviour at the same configuration
. It means that the particle's spin is parallel to the symmetry axis (positive spin) the effective potential is the same as the potential of the counterclockwise ($\mathcal{L}>0$) moving particle with antiparallel spin (negative spin) Ref. \cite{Benavides-Gallego:2018htf,Benavides-Gallego:2021lqn}.

However, the presence of electromagnetic interaction destroys the symmetry in the effective potential and the conditions for stable circular orbits. In this sense, throughout this work, we plan to study the dynamics of a spinning particle with a magnetic dipole moment around a Schwarzschild black hole immersed in an external asymptotically uniform magnetic field assuming the particle's spin, dipole moment and the external magnetic field lines are parallel 
and all of them are perpendicular to the equatorial plane.   

This paper is organized as follows: in section~\ref{section2} we present a study of the motion of test-spinning magnetized particles around magnetized Schwarzschild BH. In Section~\ref{section3} we investigate collisions of spinning magnetized particles in the spacetime of magnetized Schwarzschild black holes. We discuss neutron stars orbiting massive black holes as spinning magnetized particles in Section \ref{section4}. In the last section \ref{conclusion}, we summarize all the obtained main results.  


We use the signature $(-,+,+,+)$ for the spacetime metric and geometrized unit system $G = c = 1$. Latin indices run from $1$ to, $3,$ while Greek ones take values from $0$ to $3$.

 \section{Test spinning magnetized particle motion}\label{section2}
 
The spacetime around a Schwarzschild black hole is described in spherical coordinates, ($x^{\alpha}=\{t,r,\theta,\phi\}$) as:
\begin{eqnarray}\label{metric}
ds^2=-f(r)dt^2 + \frac{1}{f(r)}dr^2+r^2 \left(d\theta^2  + \sin^2\theta d\phi^2 \right)\ ,
\end{eqnarray}
with the radial function $f(r)= 1-{2M}/{r}$, where $M$ is the total mass of the black hole. 

\subsection{Magnetized Schwarzschild black holes}
The classical Wald solution~\cite{Wald74} for the electromagnetic four-potential of the external asymptotically uniform electromagnetic field around the Schwarzschild black hole has the form,  %
\begin{eqnarray}
A_{\varphi} = \frac{1}{2}B_0r^2\sin^2\theta \ ,
\end{eqnarray}
the asymptotic value of the magnetic field is $B_0$. The non-zero components of the electromagnetic tensor ($F_{\mu\nu}=A_{\nu,\mu}-A_{\mu,\nu}$) are
\begin{eqnarray}\label{FFFF}
F_{r \varphi}&=&B_0 r \sin^2\theta\ ,
 \\
 F_{\theta \varphi}&=&B_0r^2\sin\theta \cos\theta\ , 
\end{eqnarray}

and non-zero components of the magnetic fields around the Schwarzschild BH calculates as,

\begin{eqnarray}\label{fields}
B^{\alpha} &=& \frac{1}{2} \eta^{\alpha \beta \sigma \mu} F_{\beta \sigma} w_{\mu}\ ,
\end{eqnarray}
where, $w_{\mu}$ is the four-velocity of the proper observer, $\eta_{\alpha \beta \sigma \gamma}$, 
is the pseudo-tensorial form of the Levi-Civita symbol, and it has the following relations:  
\begin{eqnarray}
\eta_{\alpha \beta \sigma \gamma}=\sqrt{-g}\epsilon_{\alpha \beta \sigma \gamma}\, \qquad \eta^{\alpha \beta \sigma \gamma}=-\frac{1}{\sqrt{-g}}\epsilon^{\alpha \beta \sigma \gamma}\ .
\end{eqnarray}

Here $g$ is the determinant of the spacetime metric, and it is for the Schwarzschild case $g=-r^4\sin^2\theta$, and the Levi-Civita symbol $\epsilon_{0123}=1$ for the even permutations, and for odd ones $-1$.

The orthonormal components of the magnetic fields can be expressed using the electromagnetic field tensor in the following form:
\begin{equation}\label{Bi}
    B^{\hat i}=\frac{1}{2}\epsilon_{ijk}\sqrt{g_{jj}g_{kk}}F^{jk}=\frac{1}{2}\epsilon_{ijk}\sqrt{g^{jj}g^{kk}}F_{jk}
\end{equation}

Consequently, the radial and vertical components of the magnetic field measured by Zero Angular Momentum Observer (ZAMO) with the velocities $u^{\mu}_{ZAMO}=(1/\sqrt{f(r)},0,0,0)$ take the form,

\begin{equation}\label{Br}
    B^{\hat{r}}=B_0 \cos\theta, \ B^{\hat{\theta}}=\sqrt{f(r)}B_0\sin \theta .
\end{equation}

\subsection{Effective potential for a spinning magnetized particle's motion}

In fact, the effects of external magnetic fields in test-charged particles' dynamics play an important role. The motion and radiation of the charged particles around a magnetized black hole, taking into account radiation reaction forces, were extensively studied in Refs. \cite{Stuchlik2020Univ,Stuchlik2021Univ}. Also, studies of the motion of magnetized particles around black holes in the presence of external magnetic fields together with the spin effects have an interesting future in their dynamics. 

In this study, we investigate for the first time, the dynamics of the spinning test particle with a magnetic dipole moment orbiting a magnetized Schwarzschild black hole. In fact, the equations of motion of spinning particles cannot be described by the geodesic equation in general relativity due to the presence of gravitational interaction between the Riemann curvature tensor and the spin of the orbiting particle in the black hole spacetime ~\cite{Mathisson:1937zz,Papapetrou:1951pa,Corinaldesi:1951pb}.
    
Similarly, the Tulczyjew's method is an analogue method of Mathisson~\cite{tulczyjew1959motion, BWTulzcyjew1962} while Moller and others have also improved the studies defining the center of mass ~\cite{moller1949definition,beiglbock1967center,dixon1964covariant, Dixon:1970zza, Dixon:1970zz, ehlers1977dynamics}. 

Generally, the equations for the motion of particles that have spin and mass are called the MPD equations. Recently, some authors modified the MPD equations; see Refs.~\cite{Deriglazov:2017jub, Deriglazov:2018vwa} in the form,

\begin{eqnarray}\label{mpde}
\frac{Dp^\mu}{d\tau}+\frac{1}{2}R^\mu_{\nu \alpha \beta}u^\nu S^{\alpha \beta}=0 , \\
\frac{DS^{\mu \nu}}{d\tau}-p^{\mu}u^{\nu}+p^{\nu}u^{\mu}=0, 
\end{eqnarray}
where the covariant derivative reads as $D/d\tau \equiv u^\mu \nabla_\mu$ and $R^\mu_{\nu \alpha \beta}$ is the Riemann tensor. $S^{\alpha\beta}$ is the second rank antisymmetric tensor $S^{\alpha\beta}=-S^{\beta\alpha}$ describing the spin of particles as follows

\begin{eqnarray}\label{HJ2}
S^{\mu\nu} S_{\mu\nu}=2S^2=2m^2s^2\ ,
\end{eqnarray}
where the term $R^\mu_{\nu \alpha \beta}u^\nu S^{\alpha \beta}$ describes the gravitational interaction between the spacetime and spin of particles with the mass $m$. 

We also assume that a test spinning particle has a proper magnetic dipole moment together with the spin parameter, being parallel to each other, and there is an additional interaction between the dipole moment and the external magnetic field. The four-momentum taking into account the magnetic interaction has the following form \cite{deFelice,Juraeva2021EPJC,Bokhari2020PRD},

\begin{eqnarray}\label{HJ}
p^\mu p_\mu=-m^2\Big(1-\frac{1}{2m} D^{\mu \nu}F_{\mu \nu}\Big)^2\ ,
\end{eqnarray}
where $D^{\mu \nu}$ is the polarization tensor and the product $D^{\mu \nu}F_{\mu \nu}$ corresponds to magnetic interaction, and it is scalar. It is seen from the right-hand side of Eq.(\ref{HJ}) that the effective mass of the test particles in the presence of the electromagnetic field is $m_{\rm eff}=m-(1/2)D^{\alpha \beta}F_{\alpha \beta}$.
The tensors $D^{\mu \nu}$ and $S^{\mu \nu}$ are for particles with the magnetic dipole moment and spin can be expressed as,

\begin{equation}\label{Dab}
    D^{\alpha \beta}=\eta^{\alpha \beta \sigma \nu}u_{\sigma}\mu_{\nu} , \qquad  S^{\alpha \beta}=\eta^{\alpha \beta \sigma \nu}u_{\sigma}s_{\nu}.
\end{equation}



One can easily see from Eq.(\ref{Dab}) that in the case when the particle's magnetic moment and spin axes of the particles are orthonormal to the four-velocity of the particle (that means the magnetic dipole moment and spin have only $\theta$ component: $\mu^{i}=(0,\mu^{\theta},0)$ and $s^{i}=(0,s^{\theta},0)$ ) the condition must be satisfied for both spin and polarization tensors,

{\begin{equation}\label{ssc}
    S^{\alpha \beta }p_{\beta}=0\ , \qquad D^{\alpha \beta }p_{\beta}=0.
\end{equation}}

Furthermore, adding to the Tulczyjew Spin Supplementary Condition (SSC) Eq.~(\ref{ssc}), we have conserved quantities related to the space-time symmetries.  Space-time metric Eq.~(\ref{metric}) has two Killing vector fields. One generates invariant time translation $\xi^\alpha$, and the other invariant rotation $\psi^\alpha$. For this reason, we have two conserved quantities which could be found in the equation below, 

\begin{eqnarray}
    \label{s2e8}
    p^\alpha \kappa_\alpha-\frac{1}{2}S^{\alpha\beta}\nabla_\beta\kappa_\alpha = p^\alpha \kappa_\alpha-\frac{1}{2}S^{\alpha\beta}\partial_\beta\kappa_\alpha
= {const},
    \end{eqnarray}
    where $\kappa^\alpha$ is a vector related to the two Killing vector fields; i.e., $\xi^\alpha$ or $\psi^\alpha$.

The product $D^{\alpha \beta}F_{\alpha \beta}$ can be found taking into account the conditions given in Eqs.~(\ref{Dab}) and (\ref{Br}) as,
\begin{eqnarray}\label{DF1}
 D^{\mu \nu}F_{\mu \nu}=2\mu^{\hat{\alpha}}B_{\hat{\alpha}}
 =2\mu B_0 \sqrt{f(r)}\ ,
 \end{eqnarray}
where $\mu^2= \mu_{\alpha}\mu^\alpha$ is the norm of the dipole magnetic moment of magnetized particles. Here we have assumed that the direction of the magnetic dipole moment is perpendicular to the equatorial plane, it is parallel to the external magnetic field as well. We also consider the equatorial motion of the test spinning magnetized particles which makes simpler our further calculations. 

{In fact, the magnetized particle's magnetic dipole moment has interaction with only the external magnetic field that does not break the conservative quantities $p_t$ and $p_\phi$ of spinning particles, and they can be obtained from Eq.~(\ref{s2e8}) as:

\begin{eqnarray}\nonumber 
-E&=&p_t-\frac{1}{2}g_{t\alpha,\beta}S^{\alpha\beta}=p_t-\frac{1}{2}g_{tt,r}S^{tr},\\
~J&=&p_\varphi-\frac{1}{2}g_{\varphi\alpha,\beta}S^{\alpha\beta}=p_\varphi-\frac{1}{2}g_{\varphi\varphi,r}S^{\varphi r} \label{s2e10}, 
\end{eqnarray} 
where $J$ is the total angular momentum of the spinning particle, and it can be described as $J=L+S$ ($S=s m$, $L={\cal L}m$).

In fact, in a plane near the equatorial one, magnetic field lines are not perpendicular to that plane. According to our assumption, the magnetic dipole moment of the particle is perpendicular to the plane where its motion occurs. Thus, the external magnetic field and magnetic dipole moment are not parallel to each other. In other words, there is a non-zero angle between the particle's magnetic moment and the external magnetic field. It is known that the potential energy of the magnetic interaction reaches its minimum in an equilibrium state when the angle is zero. Consequently, a "non-linear" force appears that brings the magnetized particle back to its equilibrium state which causes starting "non-linear" oscillations of the magnetic dipole around the vertical axis. Then, the particle radiates electromagnetic waves which causes the particle either fall into or escape from the central object losing its energy and angular momentum. Thus, the motion of magnetized particles at a non-equatorial plane is not stable. Therefore,} we restrict our attention to the equatorial plane, where $\theta=\pi/2$.

{Using the  Eq.~(\ref{HJ2}) and Eq.~(\ref{ssc}) conditions one may get the equation below.
\begin{eqnarray}
S^{tr}=-\frac{p_\varphi s}{\sqrt{-g_{tt}g_{rr}g_{\varphi\varphi}}};\quad
S^{\varphi r}=\frac{p_t s}{\sqrt{-g_{tt}g_{rr}g_{\varphi\varphi}}};
\end{eqnarray}}

{The Eq.~(\ref{s2e10}) can be described as:
\begin{eqnarray}\nonumber 
-E&=&p_t+\frac{s~ p_{\varphi} g_{tt,r}}{2\sqrt{-g_{tt}g_{rr}g_{\varphi\varphi}}}=p_t-\frac{s}{2r}p_{\varphi}f',\\
~J&=&p_\varphi-\frac{s~p_t g_{\varphi\varphi,r}}{2\sqrt{-g_{tt}g_{rr}g_{\varphi\varphi}}}=p_{\varphi}-s ~p_t  \label{s2e12}, 
\end{eqnarray}   }
{and by solving Eq.~(\ref{s2e12}) we get:
\begin{eqnarray}
p_t=-\frac{2 E r-s J f'}{2r-s^2f'};\quad
p_\varphi=\frac{2(J r-sEr)}{2r-s^2f'} .\label{s2e14}
\end{eqnarray}     
Now, we use Eq.~(\ref{HJ}) and Eq.~(\ref{DF1}) to get the equation for the effective potential of the spinning magnetized particle.
\begin{equation}\label{s2e15}
    (p^r)^2=-g^{rr}\Big[g^{tt}p_t^2+g^{\varphi\varphi}p_\varphi^2+m^2\Big(1-\beta\sqrt{f(r)}\Big)^2\Big],
\end{equation}
where $\beta=\mu B_0/m$ is the magnetic interaction (coupling) parameter that describes the interaction between the dipole moment and the magnetic field.

\begin{equation}\label{eqmotion}
   \rho \ (u^r)^2= \alpha {\cal E}^2+\delta {\cal E} + \gamma , 
\end{equation}
where ${\cal E}$ is the specific angular momentum, and new notations are
\begin{eqnarray}
    \rho &=& (2r-s^2 f')^2, \quad  \alpha= 4(r^2-s^2f),  \\ \nonumber
    \delta &=&4s J (2f-rf') ,\\ \nonumber
    \gamma &=& -f(2r-s^2f')^2(1-\beta\sqrt{f})^2-{\cal J}^2[4f-(sf')^2]\ ,
\end{eqnarray} 
with the specific angular momentum ${\cal J}=J/m$. 

We can rewrite Eq.(\ref{eqmotion}) as,
\begin{equation}\label{pr2ex} 
(u^r)^2 = \frac{\alpha}{\rho}({\cal E}-V_+)({\cal E}-V_-).
\end{equation}

Hence, it is possible to define the effective potential for the circular motion of the spinning magnetized particles  $p^r=0$ as a solution of Eq.(\ref{pr2ex}) in the following form \cite{Toshmatov:2019bda,Toshmatov:2020wky,Benavides-Gallego:2018htf},

\begin{equation}
    V_\pm= \frac{-\delta \pm \sqrt{\delta^2-4\alpha \gamma}}{2\alpha}. 
\end{equation} 

We can define new variables as
\begin{equation}
        \label{s3Ae10}
\mathcal{J}\rightarrow\frac{\mathcal{J}}{M}=\frac{J}{m M}, \qquad s\rightarrow\frac{s}{M}=\frac{S}{m M}.
    \end{equation}


One can see from Eq. (\ref{pr2ex}) that in order to have $(u_r)^2 \geq 0$, the specific energy of the test particles has to satisfy the conditions: (i) ${\cal E}<V_{-}$ or (ii) ${\cal E}>V_{+}$. 

Hereafter, we focus on the case of spinning test particles with positive energy which coincides with the effective potential to be $V_{\rm eff} = V_+$. Here we redefine the effective potential as,

\begin{equation}\label{effpoteq}
    V_{eff}= \frac{-\delta + \sqrt{\delta^2-4\alpha \gamma}}{2\alpha}
\end{equation}

Now, we analyse the effects of spin and magnetic interactions, as well as combined effects on the effective potential for the radial motion of test spinning and magnetized particles graphically due to the complicated form of the expression for the potential.

\begin{figure*}[ht!]
\centering
    \includegraphics[scale=0.55]{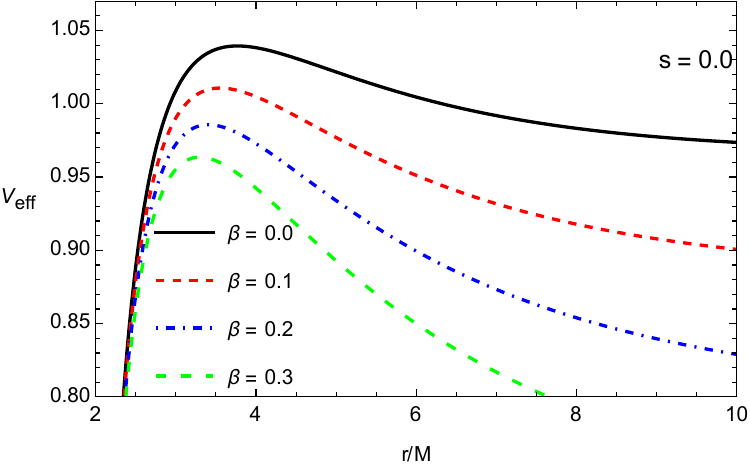}
    \includegraphics[scale=0.55]{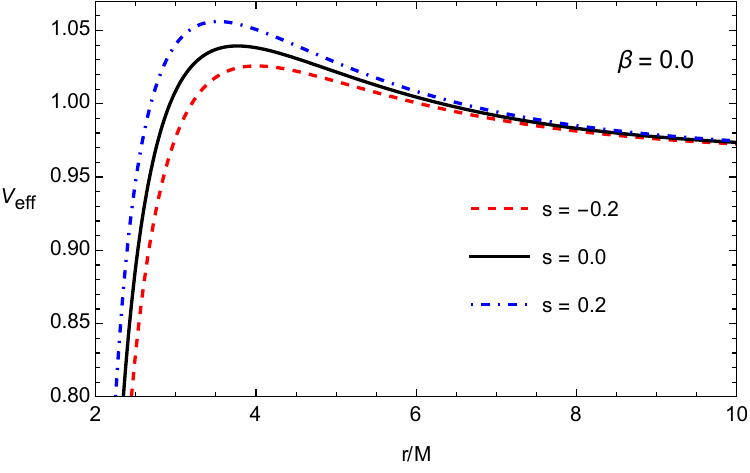}
    \includegraphics[scale=0.55]{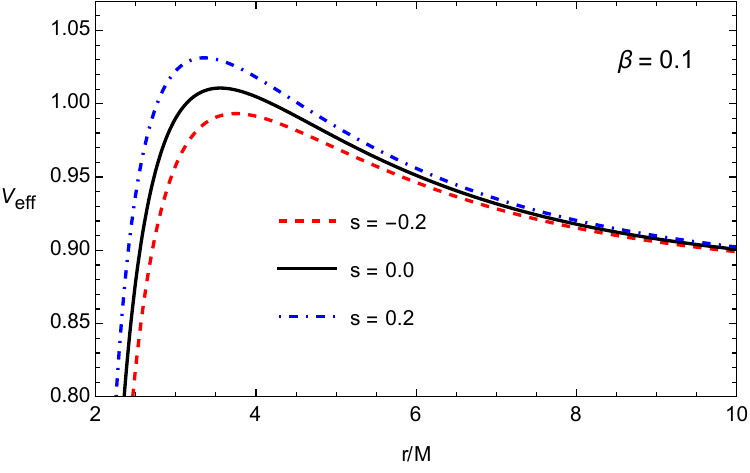}
    \includegraphics[scale=0.55]{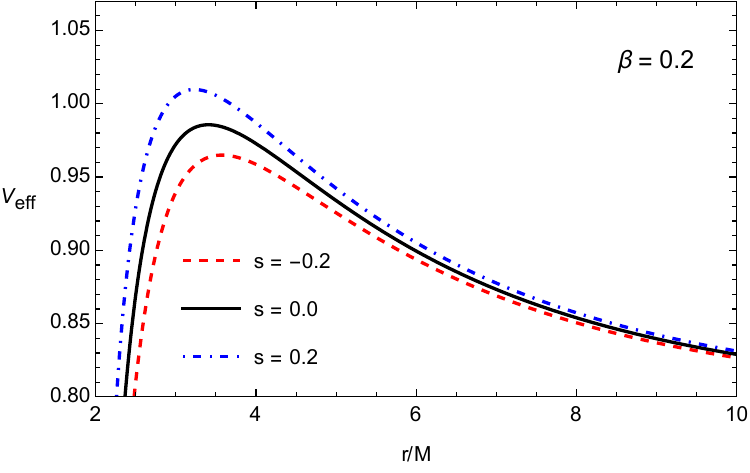}
\caption{Radial dependence of effective potential for the different values of the magnetic parameter $\beta$ and spin of the particle. In all cases ${\cal L}=4.3M$\label{vefp}}
    \end{figure*}
    
Figure \ref{vefp} displays the radial dependence of the effective potential for the radial motion of the magnetized spinning particle for various values of the magnetic parameter $\beta$ and spin of the particle $s$, while the specific angular momentum of the particle is held constant at $\mathcal{L}=4.3$.

On the other hand, Fig. \ref{vefp} demonstrates that the magnetic parameter $\beta$ has a considerable impact on the effective potential; when $\beta$ increases, the effective potential decreases, as evidenced by the left-top panel, which portrays the spinless particle to exist even at larger distances. Additionally, the other three panels portray the various values of the magnetic parameter $\beta$ with the varying $s$ spin of the particle. In this case, $s$ is chosen to be $-0.2, 0, 0.2$ to observe the impact of spin on the effective potential under different values of the magnetic parameter $\beta$. 

In the right-top panel of Fig. \ref{vefp}, the effective potential of the neutral spinning particle is plotted as a function of radial distance. From the figure, one can observe that as the spin of the particle increases, the maximum of the effective potential increases correspondingly. The same behavior can be observed in the other two panels (Fig. \ref{vefp} second row).

Now, we analyze radial profiles of ${\cal E}$ and $ {\cal L}$ of the spinning magnetized particles corresponding to circular orbits graphically, using the condition $dV_{eff}/dr=0$, due to the complicated form of the effective potential (\ref{effpoteq}). 

\begin{figure*}[ht!]
\centering
    \includegraphics[scale=0.62]{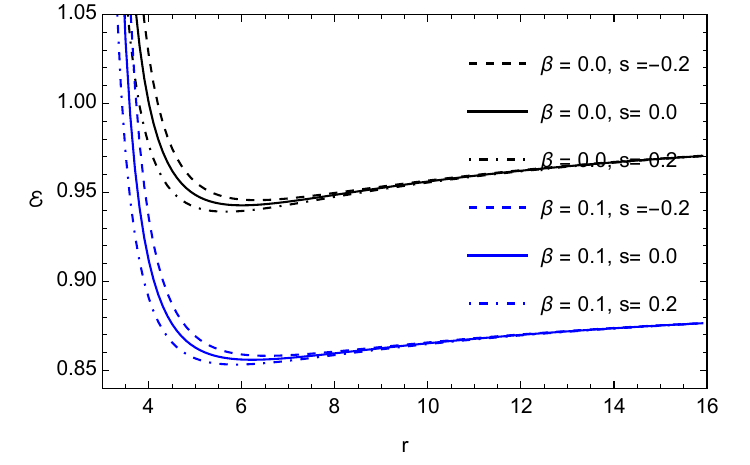}
    \includegraphics[scale=0.62]{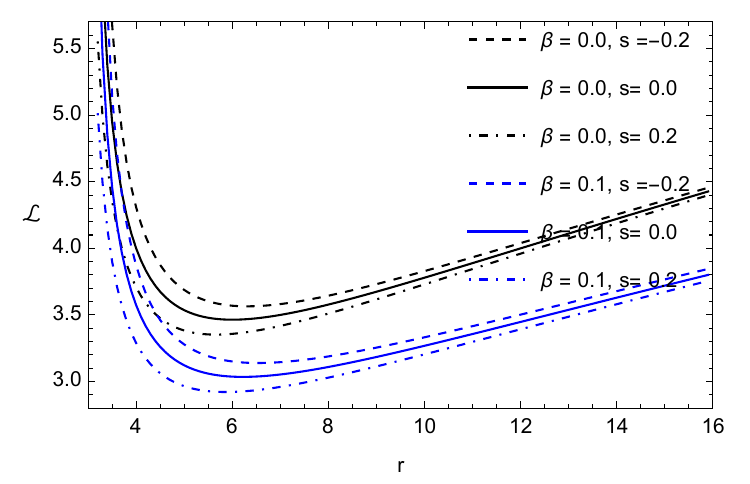}
\caption{Radial dependence of energy (left panel) and angular momentum (right panel) at circular orbits of the spinning particle for different values of the spin and magnetic parameter. Here, we have taken $M=1$.\label{lcirc1}}
    \end{figure*}

Figure \ref{lcirc1} shows radial profiles of the energy and angular momentum corresponding to circular orbits of test-spinning magnetized particles orbiting magnetized Schwarzschild black holes for different values of spin and magnetic interaction parameters. The presence of magnetic interaction sufficiently decreases both the energy and angular momentum, while the spin effects change slightly.

\subsection{Superluminal bound \label{luminal}}

In this subsection, we discuss superluminal bound. As mentioned above particle's four-velocity $u^\alpha$ and four-momentum $p^\alpha$ are not parallel in the case of the spinning particles. And a square of the four-velocity:
$$u^\alpha u_\alpha \neq -1$$
Here one can see that as the particles come close to the BH their velocities start to increase, and they may become space-like, non-physical, or in other words \textit{superluminal}. Superluminal condition is important for the motion of the spinning particles where this condition helps us to distinguish time-like particles from space-like ones, in other words, it limits the spin of the particles. Useful discussions about Superluminal bound can be found in the work of \cite{Benavides-Gallego:2021lqn,Abdulxamidov:2022ofi,Ladino:2022aja}

The condition for the spinning magnetized particle to be time-like is:
\begin{equation}
    u^\alpha u_\alpha < 0 
\end{equation}
or one can write this function as:

\begin{equation}\label{2.28}
   \frac{u^\alpha u_\alpha}{(u^t)^2}  = g_{tt} + g_{rr} \Big(\frac{dr}{dt}\Big)^2 + g_{\varphi \varphi} \Big(\frac{d\varphi}{dt}\Big)^2 < 0.
\end{equation}

With the superluminal bound holding, the components of the symmetric spin tensor $S^{\alpha\beta}$ can be computed by utilizing the method developed in Ref.~\cite{Hojman:2012me}. This method is based on the application of the MPD equations (\ref{mpde}) and the Tulczyjew-SSC. Specifically, the second MPD equation is used to calculate $DS^{tr}/d\lambda$, $DS^{t\varphi}/d\lambda$ and $DS^{r\varphi}/d\lambda$, yielding a system of equations for the non-zero components of $S^{\alpha\beta}$. This system can then be used to calculate the radial and azimuthal components of the 4-velocity vector, $u^r$ and $u^\varphi$.
    \begin{equation}
    \label{3.31}
    \begin{aligned}
    \frac{DS^{tr}}{d\lambda}&=p^tu^r-u^tp^r,\\
    \frac{DS^{t\varphi}}{d\lambda}&=p^tu^\varphi-u^tp^\varphi,\\
    \frac{DS^{r\varphi}}{d\lambda}&=p^ru^\varphi-u^rp^\varphi.
    \end{aligned}
    \end{equation}
    By making a unique gauge with the choice of $\lambda = t$, as specified in Ref.~\cite{Hojman:2012me}, the system of equations for the non-zero components of $S^{\alpha\beta}$ can be reduced to a single equation in terms of $S^{\varphi r}$ only. This is a consequence of the MPD equations, which allow for the deduction of constraints on the components of $S^{\alpha\beta}$.
    \begin{eqnarray}
        \label{3.44}
    u^r&=&\frac{{\mathcal{C}}}{{\mathcal{B}}}\frac{p_r}{p_t},\\
    u^\varphi &=&\frac{{\mathcal{A}}}{{\mathcal{B}}}\frac{p_\varphi}{p_t},
    \end{eqnarray}

    with
    \begin{eqnarray}
\label{3.45}
{\mathcal{A}}&=g^{\varphi\varphi}+R_{trrt}\left(\frac{S^{\varphi r}}{p_t}\right)^2,\\
{\mathcal{B}}&=g^{tt}+R_{\varphi rr\varphi}\left(\frac{S^{\varphi r}}{p_t}\right)^2,\\
{\mathcal{C}}&=g^{rr}+R_{\varphi tt\varphi}\left(\frac{S^{\varphi r}}{p_t}\right)^2.
    \end{eqnarray}

  By inserting Eq.~(\ref{3.44}) into the superluminal bound condition \eqref{2.28}, we can obtain a condition for the validity of the superluminal bound.
    \begin{equation}
    \label{3.47}
    g_{tt}{\mathcal{B}}^2(p_t)^2+g_{rr}{\mathcal{C}}^2(p_r)^2+g_{\varphi\varphi}{\cal A}^2(p_\varphi)^2 < 0,
    \end{equation}
    By making use of the conservation of the four-momentum, as in Eq. (\ref{HJ}), the superluminal bound condition can be simplified.
    \begin{equation}
    \label{3.48}
    \mathcal{F}=\left(\frac{p_t}{m}\right)^2\mathcal{X}+\left(\frac{p_\varphi}{m}\right)^2\mathcal{Y}-\mathcal{Z} < 0,
    \end{equation}
    with 
    \begin{equation}
    \label{3.49}
    \begin{aligned}
    \mathcal{X}&=g_{tt}{\mathcal{B}}^2-\frac{{\mathcal{C}}^2}{g_{tt}},\\
    \mathcal{Y}&=g_{\varphi\varphi}\mathcal{A}^2-\frac{\mathcal{C}^2}{g_{\varphi\varphi}},\\
    \mathcal{Z}&=g_{rr}{\mathcal{C}}^2\Big(1-\beta \sqrt{f(r)}\Big)^2.
    \end{aligned}
    \end{equation}

One can find a more detailed discussion about superluminal bound and its function in Refs. \cite{Benavides-Gallego:2021lqn,Abdulxamidov:2022ofi,Ladino:2022aja} {using the function $\mathcal{F}$, in order to have time-like particles which are moving in a circular orbit $\mathcal{F} < 0$ condition should be met.}

\subsection{Stable circular orbits}

In order to study BH, the structure of spacetime, and the accretion process, stable circular orbits are always interesting. This subsection is devoted to the innermost stable circular orbits (ISCO) of the spinning magnetized particles in the background of the Schwarzschild spacetime. There are two conditions in order to find a stable circular orbit: the radial velocity should be zero $dr/d\tau = 0$ or ${V}_{eff} = {\cal E}$ and the particle should not have radial acceleration $d^2r/d\tau^2=0$ or $d{\cal V}_{eff}/dr=0$. By using these two conditions one may find circular orbits, however, interestingly we also need to know where the inner edge of these orbits is, or just ISCO. There is another condition for the ISCO:

\begin{equation}
    \frac{d^2{V}_{eff}}{dr^2} \geq 0\ .
\end{equation}

In order to have ISCO region we need to solve a non-linear system of  equations: ${V}_{eff}={\cal E}$, $d {V}_{eff}/dr=0$ and $d^2{V}_{eff}/dr^2=0$. As these equations are non-linear we solve them numerically.

\begin{figure*}[t]
    \begin{center}
    \includegraphics[scale=0.45]{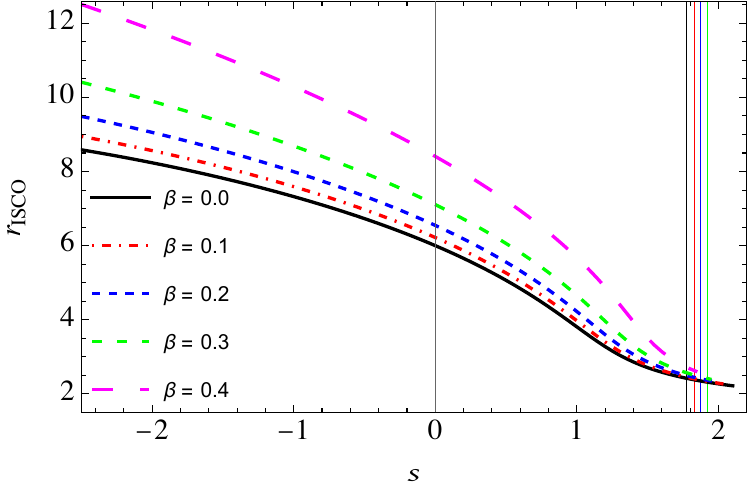}
    \includegraphics[scale=0.45]{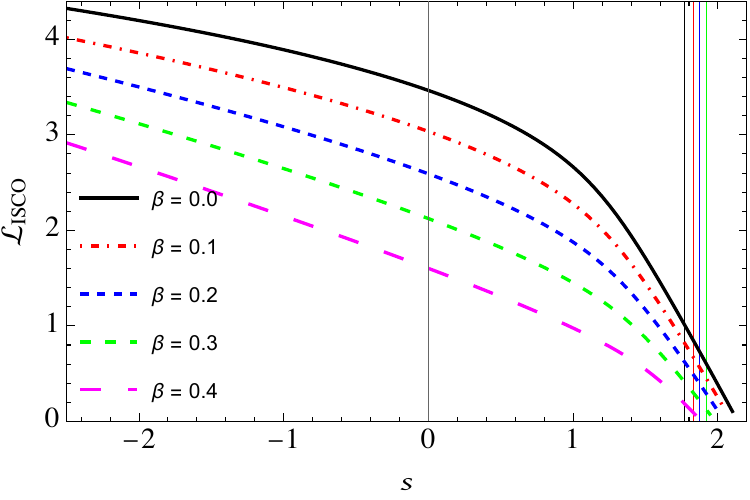}
    \includegraphics[scale=0.45]{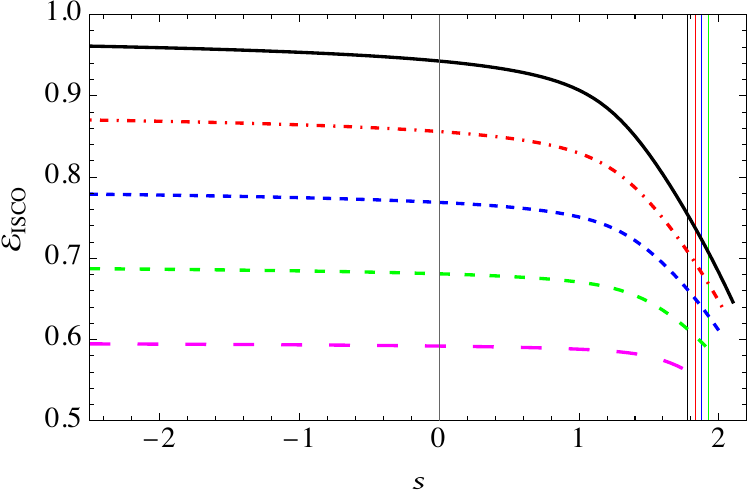}
    \end{center}
    \caption{Dependence of ISCO radius, specific angular momentum, and the specific  energy at the ISCO radius to the spin of the particle for the different values of the magnetic parameter $\beta$ (Left side plots for co-rotating case, right side for counter-rotating).\label{iscos}}
    \end{figure*}

The behaviour of $r_{ISCO}, {\cal L}_{ISCO}$ and ${\cal E}_{ISCO}$ are plotted as a function of spin $s$, in Fig. \ref{iscos}, magnetic parameter $\beta$, in Fig. \ref{iscob}, of the particle. Throughout the work, we use ${\cal L}_{ISCO}$ (${\cal J}_{ISCO}={\cal L}_{ISCO}+s$ ) notation, the reason is we want to show the effect of the spin and magnetic parameter of the particle to circular orbits.  In Fig. \ref{iscop}, relations between $r_{ISCO} $ vs ${\cal E}_{ISCO}$, $ {\cal L}_{ISCO}$ vs ${\cal E}_{ISCO}$, and $r_{ISCO} $ vs ${\cal L}_{ISCO}$ are given as a parametric plot for the values of $s\in (-2,2)$. 

In Fig. \ref{iscos}, vertical lines denote the points where Eq. (\ref{3.48}) ${\cal F} \to 0$, indicating whether the particles are time-like or space-like, as discussed in Sec. \ref{luminal}. The right side of the lines corresponds to space-like particles, which are superluminal but lack physical meaning. The left side of the lines is for the time-like particles, which are the ones of interest. One can see that magnetic parameter $\beta$ has a significant effect on the radius, specific angular momentum, and specific energy at the ISCO point. The first panel of Fig. \ref{iscos} illustrates the Innermost Stable Circular Orbit (ISCO) radius of the particles as a function of the spin of the particle. One can see that the ISCO radius increases when the magnetic parameter $\beta$ gets higher for the same spin of the particle.  Moreover, for the same magnetic parameter $\beta$, the radius, specific angular momentum, and specific energy of the particle all decrease as the spin increases. For larger values of the spin of the particle ($s\approx 2$), the ISCO radius of the particle is observed to be independent of the magnetic parameter. Fig. \ref{iscos} also shows a clear relationship between the magnetic parameter and the specific energy of the particle, indicating that the magnetic parameter has a strong effect on the specific energy of the particle. However, the specific energy of the particle decreases significantly when the spin of the particle increases from $s\approx1.5$ to $s\approx2$.

\begin{figure*}[t]
    \begin{center}
    \includegraphics[scale=0.45]{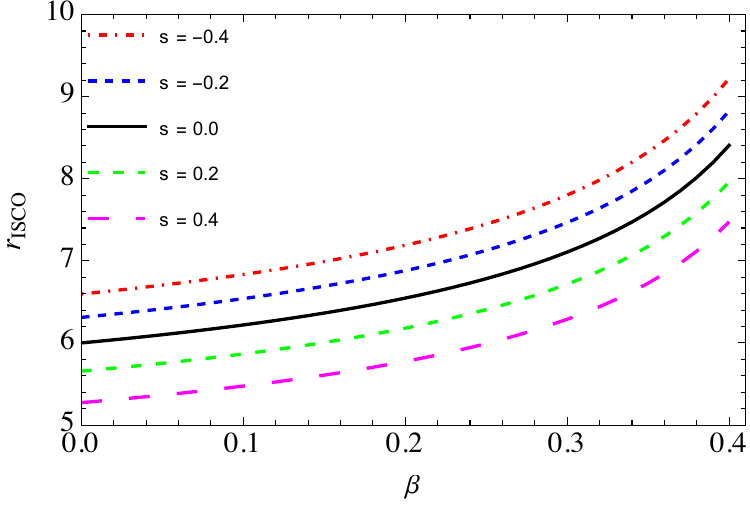}
    \includegraphics[scale=0.45]{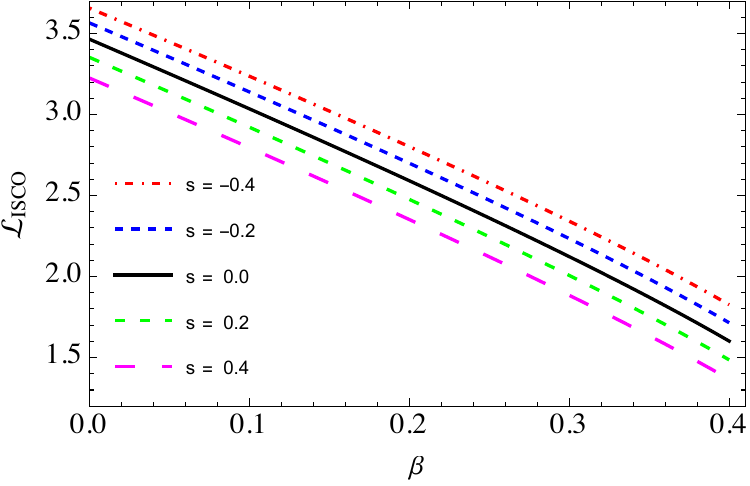}
    \includegraphics[scale=0.45]{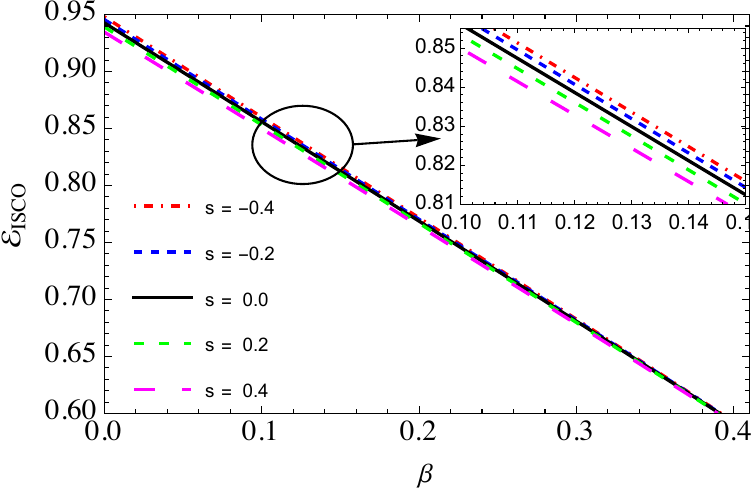}
    \end{center}
    \caption{ISCO radius, the specific energy and specific angular momentum of the particles at ISCO as a function of $\beta$ for different values of $s$.\label{iscob}}
    \end{figure*}

In Fig. \ref{iscob}, the behaviour of the radius, specific angular momentum, and specific energy of the particle at the ISCO is depicted by varying the spin ($s =-0.4; -0.2; 0.0; 0.2; 0.4$) of the particle as a function of the magnetic parameter. From the figure, one can observe a similar trend for the specific angular momentum and specific energy of the particle; both decrease with an increase in the magnetic parameter $\beta$, while the spin ($s$) remains constant. The top left panel of the figure illustrates the behaviour of the radius of the particle at the ISCO. It is observed that the radius increases at a faster rate for higher values of the magnetic parameter when the spin (s) is held constant. If the magnetic parameter $\beta$ is kept constant, then an increase in the spin $s$ of the particle will result in a decrease in the radius. Similarly, the second panel of Fig. \ref{iscob} depicts the specific angular momentum ($\mathcal{L}_{ISCO}$) of the particle at the Innermost Stable Circular Orbit (ISCO). The same pattern is observed in this panel; when the magnetic parameter $\beta$ remains constant, an increase in the spin $s$ of the particle results in a decrease in the specific angular momentum. Analogously, the decrease in specific angular momentum is inversely proportional to the increase in the magnetic parameter.  Finally, the third panel of Fig. \ref{iscob} depicts the specific energy behaviour of the particle. As depicted in the figure, the spin $s$ of the particle does not have a significant influence on the specific energy of the particle in ISCO. Furthermore, the specific energy of the particle is linearly decreasing with an increase in the magnetic parameter of the particle $\beta$.

    \begin{figure*}[t]
    \begin{center}
    \includegraphics[scale=0.45]{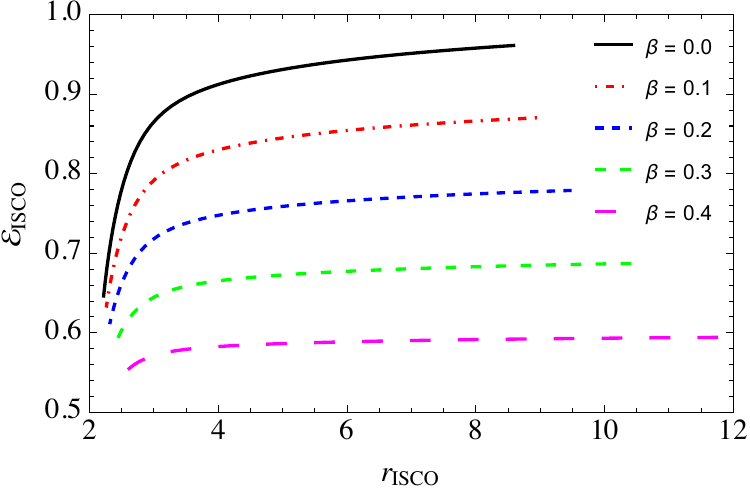}
    \includegraphics[scale=0.45]{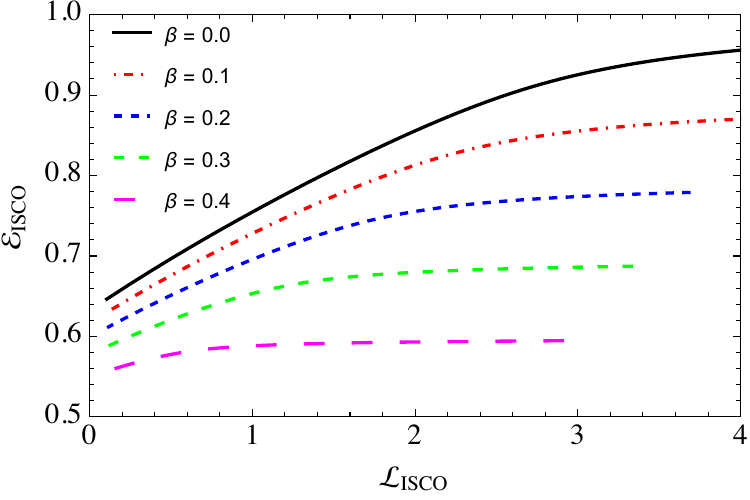}
    \includegraphics[scale=0.45]{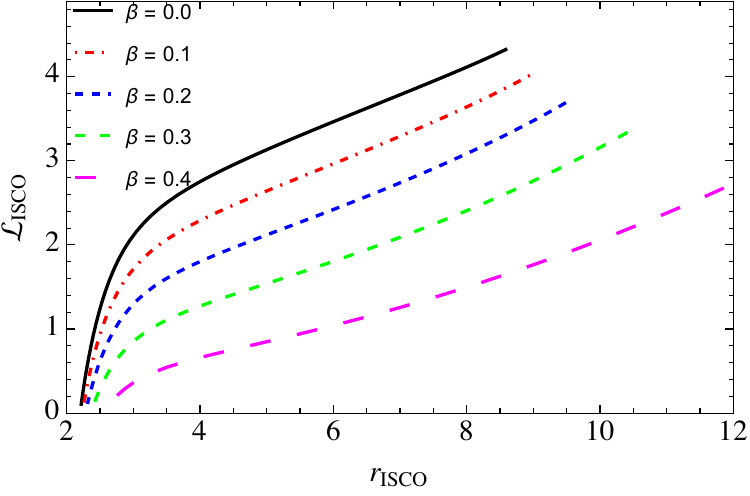}
    \end{center}
    \caption{Dependences between ISCO radius, specific energy, and specific angular momentum of the particles at ISCO for different values of $\beta$.\label{iscop}}
    \end{figure*}

Fig. \ref{iscop} shows the connection between the specific energy, angular momentum, and radius of the particle at ISCO for a range of magnetic parameter values by parameterizing the spin of the particle. The first left panel of Figure \ref{iscop} illustrates the relationship between the specific energy and radius at the ISCO for a range of values of the magnetic parameter $\beta$. Higher values of the magnetic parameter $\beta$ necessitate the spinning magnetized test particle to travel in a circular orbit with lower energy (when the radius of the circular orbit is held constant). The same behaviour for the spinning magnetized test particle can be observed in the second panel of Fig. \ref{iscop} when the angular momentum of the particle is held constant in a Schwarzschild spacetime. Finally, the last panel of Fig. \ref{iscop} reveals the amount of angular momentum needed for the spinning magnetized particle to travel in a circular orbit, where an increase in the magnetic parameter causes the angular momentum for a specific circular orbit to decrease. Furthermore, larger values of the circular orbit necessitate a greater amount of angular momentum.

\begin{figure}
    \centering
    \includegraphics[scale=0.55]{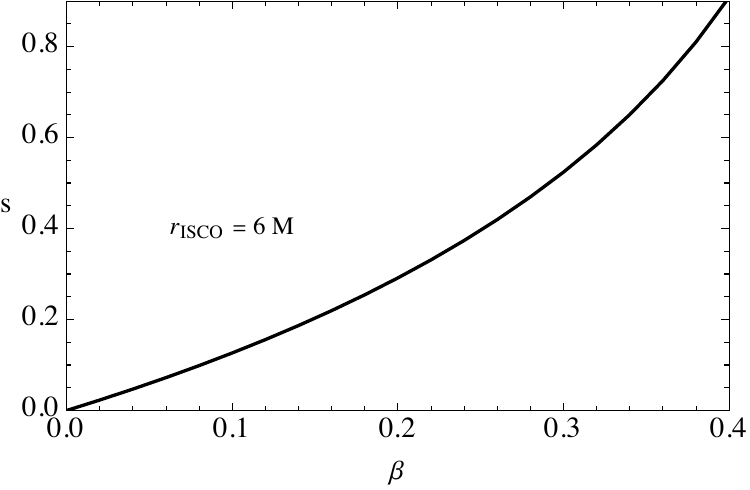}
    \caption{The relationships between the spin and magnetic parameters providing the ISCO radius as $r=6M$. \label{svb}}
\end{figure}

Finally, Figure \ref{svb} shows the correlation between the magnetic parameter and the spin of the particle for a fixed circular orbit (with $r_{ISCO}=6 M$). As depicted in the figure, the spin and magnetic parameters of the particle can mutually offset each other. Therefore, if the spinning magnetized particle is traveling with a specific amount of spin and magnetic parameter, one cannot differentiate between neutral nonspinning and magnetized spinning particles.

\section{Collisions of spinning magnetized particles}\label{section3}

For the first time, Banados-Silk-West (BSW) theoretically have been analysed the fascinating process of high-energy collisions of particles close to the black hole horizon that may be a candidate process for the energy release from the black hole \cite{Banados09}. Subsequently, numerous studies have been conducted in various settings to build on their work \cite{Abdujabbarov13a,2022Univ....8..549R,Stuchlik:2014iia}. It has been established that head-on collisions more effectively extract energy from the central black hole. 

The first energy extraction from rotating Kerr black holes is suggested in Ref.~\cite{Penrose69} where the acceleration processes become more efficient when the primary particle decays by two in the ergoregion. When ergoregion disappears, the energy released from the black hole does not extract. So, big ergoregion corresponds to big energy efficiency. Moreover, the process is more efficient in the presence of an external magnetic field and is called the magnetic Penrose process (MPP)\cite{wagh85,Dadhich18}. Protons and ions could be accelerated near magnetized supermassive black holes up to energies $10^{22}$ eV, which is explains the highest energetic protons observed in cosmic rays \cite{Tursunov2020ApJ}

This section is devoted to investigating head-on collisions and the center of mass energy of two colliding spinning magnetized particles in a Schwarzschild spacetime. It is worth noting that we have also examined the critical angular momentum of the spinning magnetized particles, which enables the particles to approach the central object from infinity.

In the following, we assume that the energy-to-mass ratio for particles arriving from infinity is equal to one for both $\mathcal{E}_1$ and $\mathcal{E}_2$. We first derive the expression for the center of mass energy of the colliding spinning magnetized particles as follows \cite{Grib11,Grib11a}

\begin{eqnarray}\nonumber 
   E_{cm}^2&=&-g^{\mu\nu}(p^{(1)}_\mu+p^{(2)}_\mu)(p^{(1)}_\nu+p^{(2)}_\nu)\\
    &=& m_1^2+m_2^2-2g^{\mu\nu}p^{(1)}_\mu p^{(2)}_\nu
\end{eqnarray}
   
Here, $p_{\mu}^{(1)}$ and $p_{\mu}^{(2)}$ are the momentum of the first and second particle, respectively, given in Eqs. (\ref{s2e14}) and (\ref{s2e15}).

\subsection{Critical angular momentum}

Before starting discussions in the center of mass energy of collisions of particles with spin and magnetic parameters, we first clarify in which values of angular momentum of particles, the particles can approach the central object from an infinite distance and the collision occurs near the horizon. To achieve this, we impose a condition on the radial motion of the particle, $\dot{r}^2\geq 0$. From Eq.(\ref{s2e15}), it can be observed that an increase in angular momentum results in a decrease in radial velocity.

This suggests that there is a critical value of angular momentum that can be obtained by solving the two equations $\dot{r}^2= 0$ and ${d\dot{r}^2/dr}=0$ simultaneously. We can now conclude that particles with higher angular momentum than the critical value cannot approach close to a black hole, and instead move at a distance from the central object.

Fig.\ref{lcrp} displays the dependence of the critical angular momentum of a spinning magnetized particle on its spin and magnetic parameter, with the left and right panels depicting the variation of the magnetic parameter and spin respectively. The left panel of Fig. \ref{lcrp} shows that an increase in the magnetic parameter leads to an increase in the angular momentum. For a fixed spin, the difference in angular momentum is greater for a negative spin than for a positive spin of the particle in Schwarzschild spacetime. For a spin of the particle approximately $s\approx 1.8$, the effect of the magnetic parameter $\beta$ on the angular momentum is negligible. Moreover, there is a maximum angular momentum for certain spin values. As the spin of the particle increases, the angular momentum also increases until a maximum value is reached at approximately $s=0.8$, after which it decreases. The right panel of Fig. \ref{lcrp} depicts the critical angular momentum versus magnetic parameter for varying spin values of $s=-0.2, 0, 0.2$. The plot shows that the critical angular momentum is approximately linearly proportional to the magnetic parameter. Additionally, an increase in the spin $s$ of the particle leads to an increase in the angular momentum for a fixed value of the magnetic parameter $\beta$.

\begin{figure*}[t]
    \begin{center}
    $\begin{array}{cc}
    \includegraphics[scale=0.55]{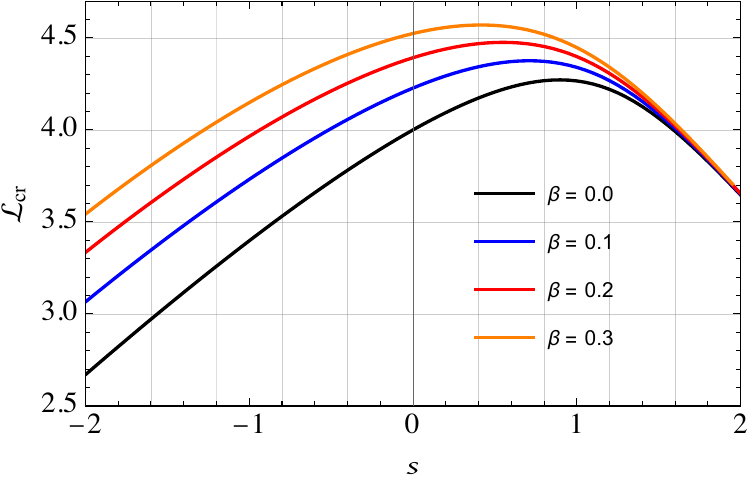}&
    \includegraphics[scale=0.55]{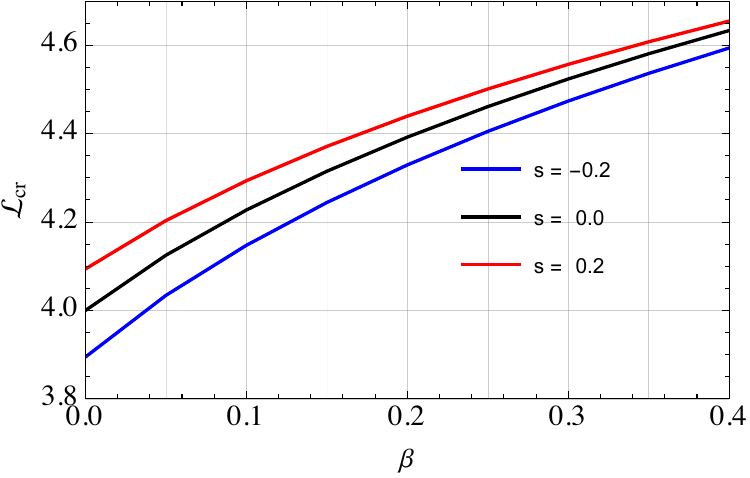}
    \end{array}$
    \end{center}
    \caption{Dependence of the critical values of the specific angular momentum for the spin ($s$) and magnetic parameter ($\beta$) of the particle respectively. 
    \label{lcrp}}
    \end{figure*}

\subsection{Collisions of spinning magnetized particles}

Here, we investigate the center of mass energy of two colliding spinning and magnetized particles near a Schwarzschild black hole.

We assume that the particles have equal mass such that $m_1=m_2=m$, and the expressions of $p_t, p_{\varphi},$ and $p_r$ are given in Eqs.(\ref{s2e12})-(\ref{s2e15}).


We can now derive an expression to calculate the center-of-mass energy of colliding particles as follows:

\begin{eqnarray}\label{ecmps}
   {\cal E}_{cm}&=&\frac{E_{cm}}{2m^2}\\ \nonumber &=&1-g^{tt}p^{(1)}_t p^{(2)}_t-g^{rr}p^{(1)}_r p^{(2)}_r-g^{\varphi \varphi}p^{(1)}_{\varphi} p^{(2)}_{\varphi}
\end{eqnarray}

Applying Eq.(\ref{ecmps}) to the expression above yields an expression for the center of mass of the colliding spinning magnetized particles.

\begin{figure*}[t]
    \includegraphics[scale=0.45]{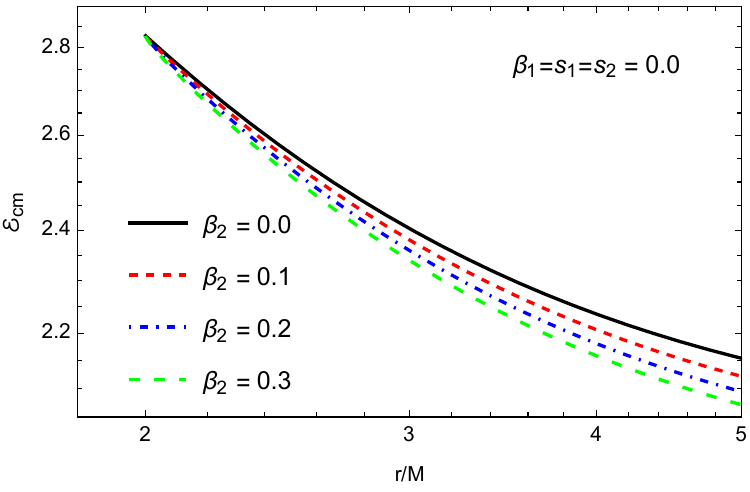}
    \includegraphics[scale=0.45]{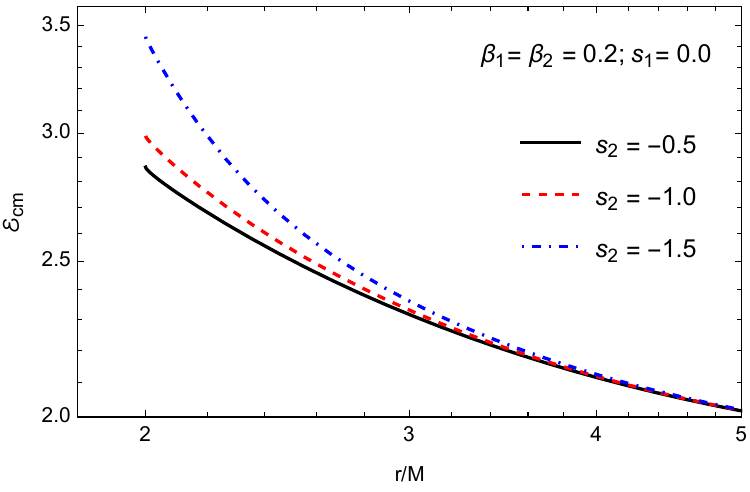}
    \includegraphics[scale=0.45]{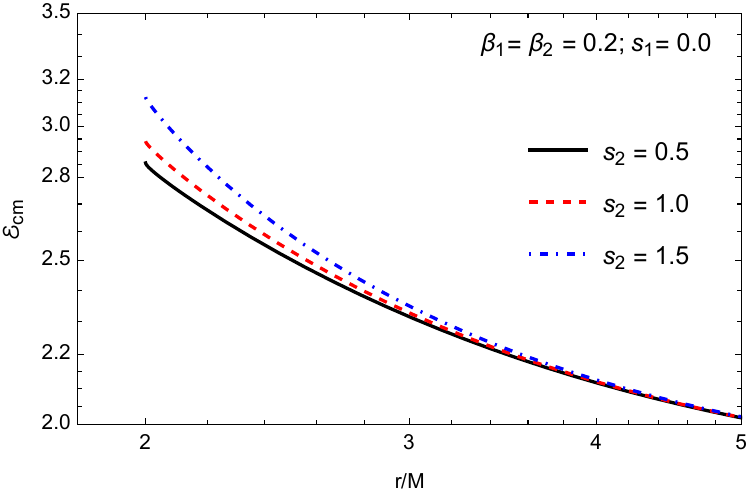}
    \includegraphics[scale=0.45]{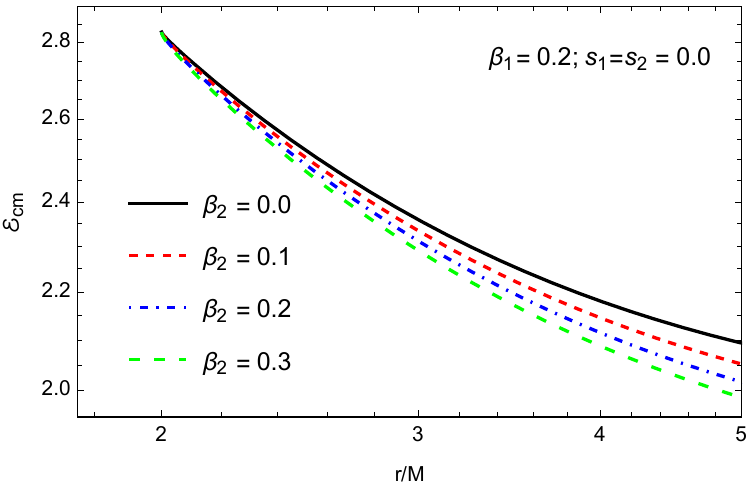}
   \includegraphics[scale=0.45]{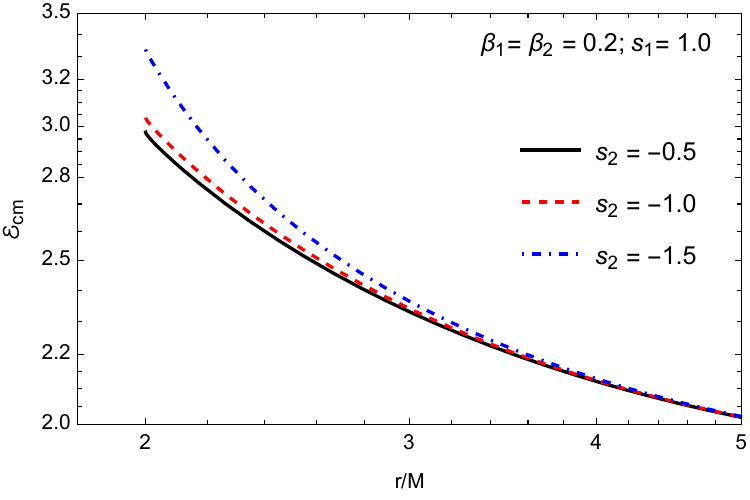}
    \includegraphics[scale=0.45]{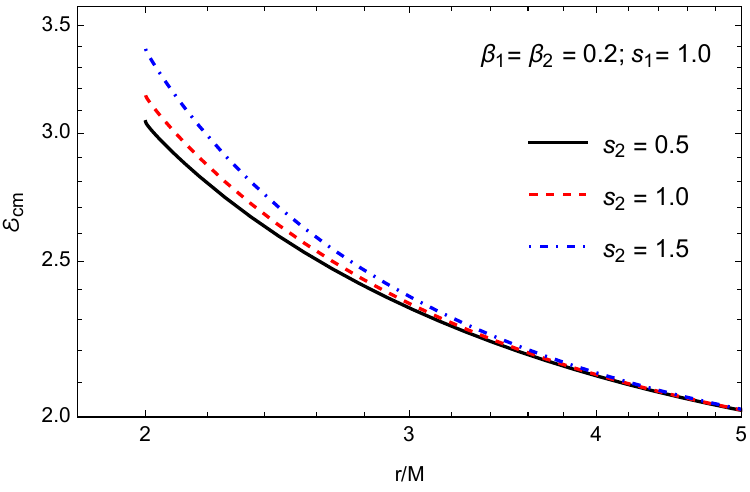}
    \caption{Radial dependence of the center of mass energy of collisions of magnetized spinning particles motion with different $s$ and $\beta$ parameters}.\label{ecmp}
    \end{figure*}

Fig. \ref{ecmp} illustrates the radial dependence of the center of mass energy $\mathcal{E}_{cm}$ of two colliding spinning magnetized particles for various values of the spin $s$ and magnetic parameter $\beta$. It consists of six panels In the top row panel, we consider one of the particles to be spinless, and in the bottom panel $\beta_1=0$. It is observed that $\beta$ decreases the energy. The top-left row is for collisions of test neutral and magnetized particles, while at the bottom-left panel, we show radial profiles of the center of mass energy of two magnetized (non-spinning) particle collisions.

The middle and right columns are for the negative and positive spinning particles' collisions, respectively. In the top row of the columns, the first particle's spin is zero while the bottom row is for the particles with spin $s_2=1$.  
One can see from the figure that the center of mass energy $\mathcal{E}_{cm}$ decreases (increases) at $s>0$ ($s<0$) cases. Moreover, the effect of $s$ is sufficiently near the horizon while $\beta$ influence on ${\cal E}_{cm}$ is visible far from the object.

\section{Astrophysical relevant spinning magnetized objects  }\label{section4}

\subsection{Neutron stars as spinning magnetized objects}

In fact, neutron stars with the magnetic dipole moment $\mu=(1/2)B_{s} R^3$, are rotating magnetized objects in the universe, where $B_{s}$ is the surface magnetic field. Now, here we simply estimate the spin parameter for the objects with the rotational angular momentum $S=I\Omega$, where $\Omega=2\pi/P$ and $P$ is the rotational period of the neutron star.

As mentioned in Eq.(\ref{s3Ae10}) that the dimensionless spin parameter for NSs can be calculated as
\begin{equation}
    s_{\rm NS}=\frac{I\Omega}{m_{\rm NS}M_{BH}}\ ,
\end{equation}
where $I$ is the inertia momentum of the NS. In Newtonian gravity, spherical massive bodies, NSs as well, have inertia momenta $I=(2/5)m_{NS}R^2$. However, the inertia momentum of an object changes due to the strong gravity and matter inside it.
%
In our previous work \cite{Turimov2021PhRvD}, we have estimated the moment of inertia of neutron stars assuming it consists of uniform density matter $\rho=const$, showing that the inertia momenta in GR do not change much even for most massive NSs, the difference is about 3\%.

Now, we evaluate the spin parameter for a neutron star orbiting an intermediate-mass black hole as,
\begin{equation}
    s_{\rm NS}\simeq 6\times 10^{-3}\frac{R_6^2}{M_4P_{ms}}
\end{equation}
where $R_6=R_{\rm NS}/(10^{6} \rm cm) $ is a normalized radius of the star to $10^6$ cm, {$P_{ms}$ is the rotational period of the NS expressed in milliseconds} and $M_{4}=M_{\rm BH}/(10^4 M_\odot) $ is the mass of the central black hole. Also, the magnetic coupling parameter is,
\begin{equation}
        \beta_{\rm NS}\simeq3\times10^{-3}\frac{B_{12}R^3_6B_1}{M_{14}}\,
\end{equation}
where $B_{12}=B_{\rm NS}/(10^{12 }\rm G)$ and $B_{1}=B_{\rm extr}/(10 \rm G)$ are the dimensionless surface magnetic fields of the neutron star and external local magnetic field around the central black hole, respectively. The mass of the star is also given in the normalized form $M_{14}=m_{\rm NS}/(1.4 M_\odot)$.


In fact, when $s\ll\beta$ the particle can be a magnetized particle, or when $s \gg \beta$ it is spinning. So, we are interested in the magnetic field value where the magnetic interaction parameter $\beta$ is of the same order as the spin parameter $s\approx \beta$ that satisfies the particle as a candidate for a test spin magnetized one. 
For this case, we let the mass of the NS be $1.4M_\odot$, its radius about 10 km, and the surface magnetic field $\sim 10^{12}$ G.
Our rough estimations show that the external magnetic field has to be bigger than about 20 Gauss to provide the same order of magnetic interaction with the spin-curvature interaction ($s=\beta=0.006$).

\subsection{Rotating stellar mass black holes as spinning magnetized objects}

Moreover, one can treat stellar-mass black holes with the ionized accretion disc generating a dipolar magnetic field around the black hole as test spinning magnetized particles orbiting massive black holes.

Stellar-mass black holes with ionized accretion discs can have a dipole-like magnetic field generated by the current loop. Authors of Ref. \cite{Piotrovich2010arXiv} have shown that the magnetic field value can be of the order of $10^8$ Gauss around the stellar-mass black holes and $10^4$ Gauss around supermassive black holes. The magnetic dipole moment of the current loop around a stellar-mass black hole is { $\mu_{cl}=(1/2)B_{cl}r_{cl}^3$, where $B_{cl}$ is the magnetic field generated by the current loop. Here, we assume the orbits of the current loop is located near ISCO. In order to estimate the magnetic parameter for the magnetized stellar mass black hole with the Kerr parameter $a_*$ (where $a_*=a/m_{SBH}$ is dimensionless Kerr parameter), we first calculate the radius of the current loop using the expression for the ISCO radius
\begin{eqnarray}\nonumber
R_{\rm ISCO}= 3 + Z_2 \pm \sqrt{(3- Z_1)(3+ Z_1 +2 Z_2 )} \ ,
\end{eqnarray}
where $+$ and $-$ signs stand for retrograde and prograde orbits/loops, respectively, and 
\begin{eqnarray} \nonumber
Z_1 &  = & 
1+\left( \sqrt[3]{1+a_*}+ \sqrt[3]{1-a_*} \right) 
\sqrt[3]{1-a^2_*} \ ,
\\ \nonumber
Z_2 & = & \sqrt{3 a^2_* + Z_1^2} \ .
\end{eqnarray}
The magnetic parameter for the fixed values of the rotation parameter of the stellar black hole  $a=0.5m_{SBH}$ and the loop magnetic field (see, \cite{Piotrovich2010arXiv}) $B_{cl}=10^8$ G is
}
\begin{equation}\label{betasmbh}
    \beta \simeq 3.6\times 10^{-3}\frac{B_8B_3}{M_1}\ ,
\end{equation}
where $M_{1}=m_{SBH}/10M_\odot$ is the mass of the stellar mass black hole normalized to 10$M_\odot$. $B_{8}=B_{cl}/(10^{8}\rm G)$ and $B_{3}=B_{\rm extr}/(10^3 \rm G)$.

The spin parameter $s$ for the system consisting of a dipolar magnetized stellar-mass black hole and the massive central black hole is 
\begin{equation}\label{ssmbh}
    s\simeq 10^{-3}a_* \frac{M_{1}}{M_4}\ .
\end{equation}  

{Now, it is possible to estimate values of the external magnetic field that provide the magnetic interaction to be comparable with the spin parameter for the particle (rotating stellar mass black hole) using Eqs.~(\ref{betasmbh}) and (\ref{ssmbh}). Our calculations have shown that the magnetic field have to be greater than $B_{extr}\simeq 280 G$ for $M_{BH}=10^4 M_\odot$ for the black hole with the mass $m_{SBH}=10M_\odot$ when the magnetic field of the current loop is in the order of $10^8\ G$.}
\begin{figure}
    \centering
    \includegraphics[scale=0.6]{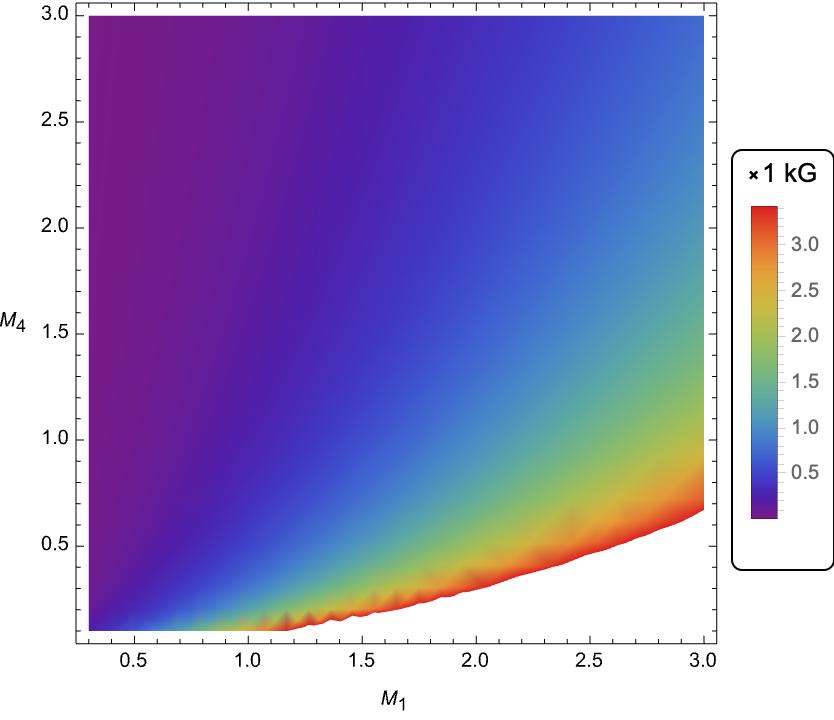}
    \includegraphics[scale=0.6]{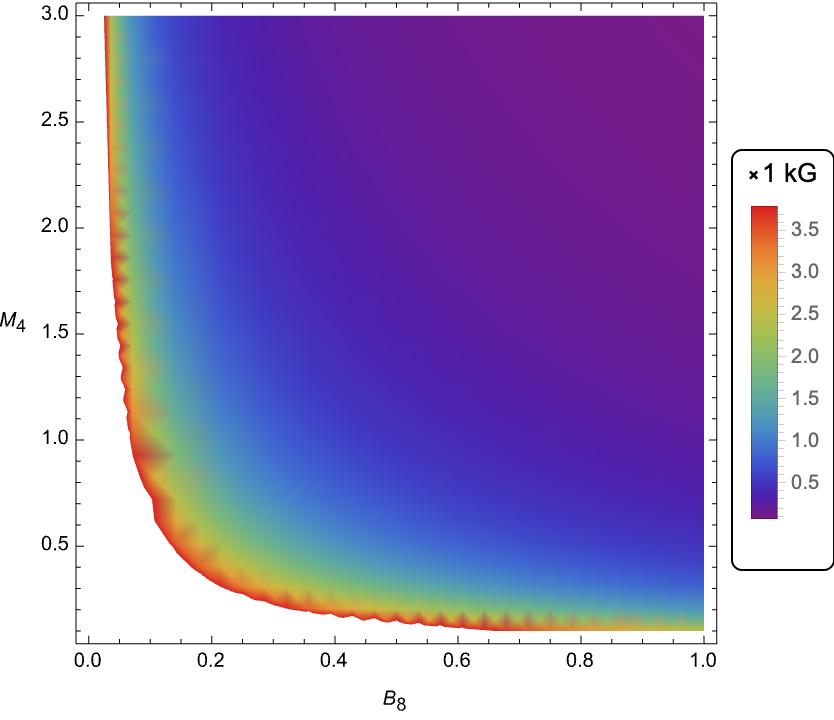}
    \caption{Densityplot of a $B_3$ as parameter of $M_1, M_4, a_*$, and $B_8$. And we have chosen $a_*=0.9$ in both panels.  \label{dp1}}
\end{figure}

{We performed a comparison between the particle spin $s$ in Eq.(\ref{ssmbh}) and the magnetic parameter $\beta$ in Eq.(\ref{betasmbh}), which allowed us to derive as a function of $B_3$. This function was then plotted and presented in Fig.\ref{dp1}. In the top panel of Fig.\ref{dp1}, where $B_8 =1 ~(\sim 10^8  \text{G})$ and $a_* = 0.9$, we observe the relationship between $M_4$ and $M_1$, with the right colorbar indicating the corresponding values of $B_3$. Furthermore, the bottom panel of Fig.~\ref{dp1} illustrates the relationship between $M_4$ and $B_8$, with $M_1$ fixed at $1$.
\\
This density plot, displaying the function $B_3$, provides insights into the interplay between the magnetic parameter, particle spin, and the masses $M_4$ and $M_1$. By examining the color-coded values of $B_3$ and the corresponding parameters, we gain a deeper understanding of their relationships and their influence on the dynamics of the system. }

\section{Conclusion}\label{conclusion}
In this paper, we have studied the dynamics of spinning magnetized particles around a Schwarzschild black hole immersed in an externally asymptotically uniform magnetic field. Here we have used the Mathisson-Papapetrou-Dixon equation for the motion of spinning test particles and the modified Hamilton-Jacobi equation, which are taken into account the gravitational interaction between the spin of the particles and magnetic interaction between the external magnetic field and magnetic dipole moment of the particles, respectively. 

We have derived the effective potential of test spinning magnetized particles' motion around the magnetized black hole. It is obtained that the effective potential decreases due to an increase in the magnetic interaction. Similarly, in the presence of negative values of the spin parameter, the effective potential decreases with a smaller effect than the effect of $\beta$. At $s>0$ the potential increases. 

We have also studied the effect of spin and magnetic interaction on innermost stable circular orbits. It is obtained that ISCOs go far from the black hole in the presence of magnetic interaction due to the repulsive behavior of Lorentz forces. The increase of the positive spin of the particle also increases the ISCO radius due to the increase of total angular momentum, while the negative one decreases. Moreover, we have also considered the case that the spinning magnetized particle's ISCO radius is the same as the neutral one's being equal to 6$M$. That means positive spin and magnetic field effects have opposite behavior and may compensate for each other with the relation given in Fig.\ref{svb}. 

We have also analysed, the energy and angular momentum of the particles at ISCO by varying $\beta$ and $s$ parameters and show that an increase of $\beta$ causes the decrease of both the angular momentum and energy at the particle's ISCO corresponding to the value of the parameter $\beta$. However, the spin parameter slightly changes the energy and angular momentum. Relationships between the ISCO radius and the energy \& angular momentum of the particles at ISCO. It is also observed that as $\beta$ increases both the energy and angular momentum for the fixed ISCO decrease. Similarly, in the presence of $\beta$ the energy of particles at ISCO decrease for the corresponding values of angular momentum of the particle at ISCO.   

We have also studied the superluminal motion of spinning magnetized particles. As shown in Fig. \ref{iscos}, the spin axis is divided into two parts by vertical colored lines, representing the time and space-like particles, respectively. The results indicate that an increase in the magnetic parameter $\beta$ leads to increasing the spin limit of the spinning particles. It means that the Lorentz force and spin-curvature interactions compensate for each other.

We have also investigated collisions of the spinning magnetized particles and calculated the critical values of the angular momentum of the particles at which the particles can collide and the center of mass energy of the collisions. It is shown that the critic value increase with the increase of the external magnetic field or magnetic moment of the particles, while, it increases with the increase of the spin of the particle, reaches its maximum, and decreases back again. However, at higher values of the spin near $s=2$, the effects of the magnetic interaction on the critic angular momentum disappear, which means that near $s=2$ the spin interaction becomes much higher than the magnetic one.  

Studies of the center of energy of the collisions have shown that the presence of positive spin and magnetic dipole moment of the colliding particles causes decreasing the energy, while negative spin causes increasing it.

Finally, we evaluated and compared the spin and magnetic interactions in the circular stable orbits of neutron stars and white dwarfs around massive black holes, considering them a test of spinning and magnetized particles.  It is found that magnetic interaction effects are much larger than the spin ones even in the presence of external magnetic fields of about mGs. Our estimations have shown that in the case of typical neutron stars orbiting intermediate-mass black holes, the external magnetic field has to be greater than 20 Gs to support a similar order of magnetic interaction with the spin interaction ($s=\beta=0.006$). {Also, the lower limit for the external magnetic field is obtained as 280 Gausses for comparable values of the spin and magnetic interaction parameters where a stellar-mass black hole is treated as a spinning magnetized object.} 

In our future studies, we plan to extend this work by investigating the dynamics of spinning magnetized particles in the close environment of magnetized rotating black holes and magnetically charged black holes in the framework of various gravity theories.

\section{acknowledgement}

This research is supported by Grant No. FA-F-2021-510 of the Uzbekistan Agency for Innovative Development. F.A., J.R., and A.A. acknowledge the ERASMUS+ ICM project for supporting their stay at the Silesian University in Opava.

\bibliographystyle{apsrev4-1}
\bibliography{references,reference2}

\end{document}